\documentclass[11pt,letterpaper]{article}

\usepackage[T1]{fontenc}
\usepackage{lmodern}

\usepackage[utf8]{inputenc} % allow utf-8 input
\usepackage[T1]{fontenc}    % use 8-bit T1 fonts
\usepackage[hidelinks]{hyperref}       % hyperlinks
\usepackage{url}            % simple URL typesetting
\usepackage{booktabs}       % professional-quality tables
\usepackage{amsfonts}       % blackboard math symbols
\usepackage{nicefrac}       % compact symbols for 1/2, etc.
\usepackage{microtype}      % microtypography

\usepackage{notoccite}
\usepackage{graphics} % for pdf, bitmapped graphics files
\usepackage{amsmath} % assumes amsmath package installed
\usepackage{amssymb}  % assumes amsmath package installed
\usepackage{color,xspace}
\usepackage[hidelinks]{hyperref}
\usepackage{graphicx}
\usepackage{cite}
\usepackage{subcaption}

\usepackage{algcompatible}
\usepackage{algorithm}
\usepackage[noend]{algpseudocode}
\usepackage{cases}
\usepackage{mdframed}

\usepackage[margin=1.0in]{geometry}

\usepackage{amsthm}

\newtheorem{theorem}{Theorem}
\newtheorem{lemma}{Lemma}

\newtheorem{example}{Example}

\newtheorem{definition}{Definition}

\title{\LARGE \bf One-Shot Reachability Analysis of Neural Network Dynamical Systems
}

\date{}

\author{Shaoru Chen, Victor M. Preciado, Mahyar Fazlyab
\thanks{Shaoru Chen and Victor M. Preciado are with the Department of Electrical and Systems Engineering, University of Pennsylvania. Email: {\tt\small \{srchen, preciado\}@seas.upenn.edu}. Mahyar Fazlyab is with the Mathematical Institute for Data Science, Johns Hopkins University. Email: {\tt\small mahyarfazlyab@jhu.edu}. } %
}

\begin{document}

\maketitle
\thispagestyle{empty}
\pagestyle{empty}

\begin{abstract}
	% abstract
The arising application of neural networks (NN) in robotic systems has driven the development of safety verification methods for neural network dynamical systems (NNDS). Recursive techniques for reachability analysis of dynamical systems in closed-loop with a NN controller, planner or perception can over-approximate the reachable sets of the NNDS by bounding the outputs of the NN and propagating these NN output bounds forward. However, this \emph{recursive} reachability analysis may suffer from compounding errors, rapidly becoming overly conservative over a longer horizon. In this work, we prove that an alternative \emph{one-shot} reachability analysis framework which directly verifies the unrolled NNDS can significantly mitigate the compounding errors {for a general class of NN verification methods built on layerwise abstraction. Our analysis is motivated by the fact that certain NN verification methods give rise to looser bounds when applied in one-shot than when applied recursively.} In our analysis, we characterize the performance gap between the recursive and one-shot frameworks for NNDS with general computational graphs. The applicability of one-shot analysis is demonstrated through numerical examples on a cart-pole system.

% and propose an efficient method that bounds the reachable sets with adaptive polytopes such that this performance gap is minimized
\end{abstract}

\section{Introduction}
% introduction
\label{sec:introduction}
Robotic systems embedding learning-enabled modules such as neural network (NN) controllers, planners, or perception have achieved state-of-the-art performances in various complex tasks and are becoming increasingly popular. However, such neural network dynamical systems (NNDS) lack formal safety guarantees and are prone to failures due to the fragility of NNs to adversarial attacks or random input perturbation~\cite{goodfellow2014explaining, huang2017adversarial}, which significantly limits their deployment in safety-critical applications such as autonomous driving. Verifying the safety of NNDS before deployment provides a promising solution, but handling the large scale and high complexity of NNs in verification is challenging.

\begin{figure}
	\centering
\includegraphics[width = 0.8 \columnwidth]{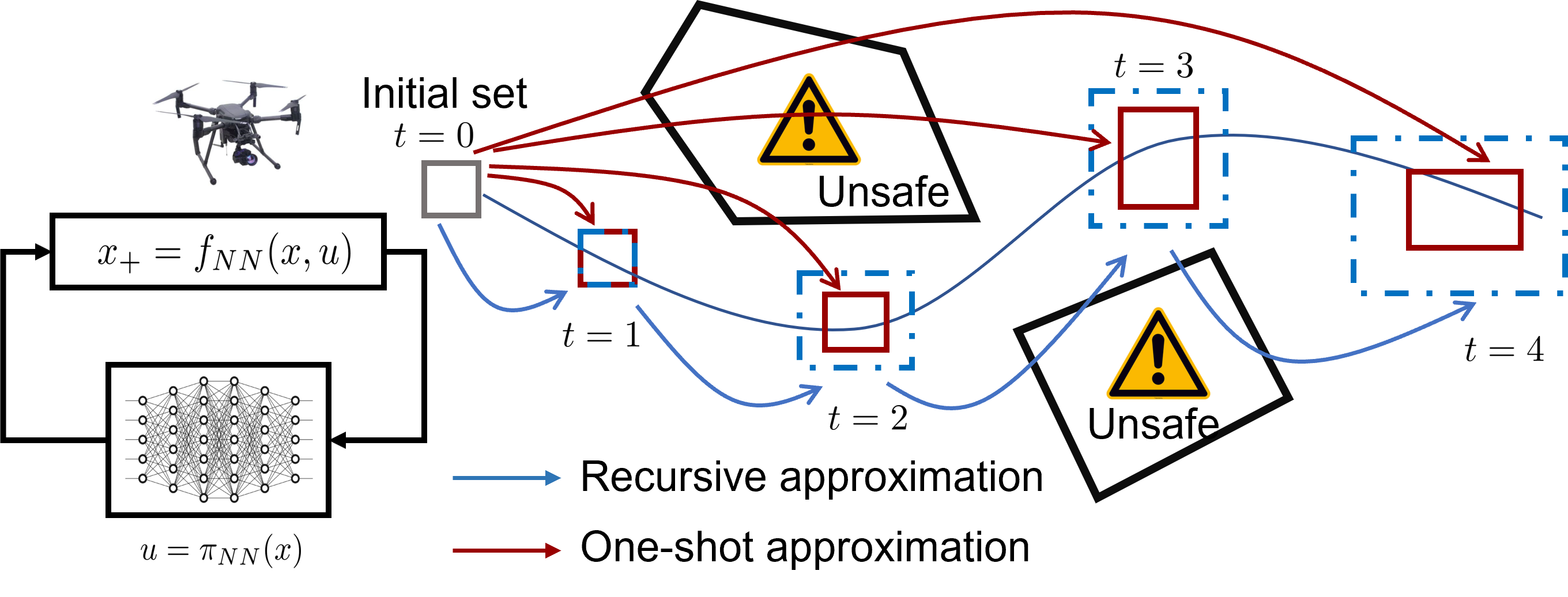}
\caption{Reachable set over-approximations of a NN dynamical system can be obtained by applying NN verification methods recursively (blue arrows) or in one-shot (red arrows). In the one-shot analysis, unrolled NN dynamics over multiple time steps is considered. In this work, we investigate in what cases the one-shot analysis generates tighter bounds than the recursive counterpart.}
\label{fig:illustration}
\end{figure}

Neural network verification consists in certifying that the output of a NN satisfies certain properties given a bounded set of inputs. A rich body of works has focused on developing specialized solvers~\cite{tjeng2017evaluating, ehlers2017formal, wong2018provable, zhang2018efficient, dvijotham2018dual, bunel2018unified, singh2019beyond, raghunathan2018semidefinite, fazlyab2020safety} for NN verification. Although NN verification methods only consider NNs in isolation, they can be conveniently combined with existing reachability analysis tools to certify properties of NNDS~\cite{akintunde2018reachability, huang2019reachnn, kochdumper2022open} over a finite horizon. For a discrete-time NNDS, NN verification methods can readily compute an over-approximation of the one-step reachable set given a bounded input set. Then, applying such one-step over-approximation method recursively for $t= 0, 1, \cdots, T$ leads to a bounding tube of the NNDS trajectories (blue boxes in Fig.~\ref{fig:illustration}). 

{We denote the above methodology as the \emph{recursive reachability analysis} framework~\cite{zhang2022safety, everett2021efficient, hu2020reach}, and compare it with an alternative framework that computes over-approximations of the reachable sets in one shot~\cite{sidrane2022overt, karg2020stability}}. In \emph{one-shot reachability analysis}, the reachable set of NNDS at time $t$ is bounded by directly applying NN verification methods on the unrolled NN dynamics for $t$ steps (red boxes in Fig.~\ref{fig:illustration}). While one may observe that the one-shot analysis generates tighter bounds than the recursive counterpart due to the use of iterated dynamics~\cite{sidrane2022overt}, this property does not hold for general NN verification tools. Correspondingly, our contributions are:
\begin{itemize}
\item We provide a counter-example where one-shot analysis results in worse bounds than the recursive framework, highlighting that the one-shot analysis does not always lead to tighter bounds.
\item We formally prove conditions under which the one-shot framework provides tighter bounds compared with the recursive framework for a general class of NN verification methods.
\item Our analysis applies to NNs with general architectures and allows us to consider disturbances in the computation of the reachable set over-approximations. Numerical examples are provided to demonstrate the applicability of the one-shot framework.
\end{itemize}

\subsection{Related works}
\paragraph{NN verification:}
NN verification methods analyze NNs in isolation and can certify the worst-case performance of NNs under bounded input perturbations. A rich set of tools with distinct tightness/complexity trade-offs are developed, including those based on mixed-integer programming (MIP)~\cite{tjeng2017evaluating}, semidefinite programming (SDP)~\cite{raghunathan2018semidefinite, fazlyab2020safety}, linear programming (LP)~\cite{wong2018provable}, and linear bounds propagation~\cite{weng2018towards, zhang2018efficient, xu2020automatic}. Particularly, \cite{xu2020automatic} considers NNs with general architectures. Tightness comparison of different convex relaxation-based NN verifiers is discussed in~\cite{salman2019convex, de2021improved} from the dual perspective. 

\paragraph{Reachability analysis of NNDS:}
Verifying a closed-loop dynamics with a NN controller often involves first bounding the set of control inputs (e.g., by polytopes~\cite{xiang2018reachable}, linear bounds~\cite{everett2021reachability}, star sets~\cite{tran2019safety}, polynomial zonotopes~\cite{kochdumper2022open}, hybrid automaton~\cite{ivanov2019verisig}, Taylor models~\cite{schilling2022verification}) in each control cycle, and then estimating the reachable set of the states. Several works~\cite{huang2019reachnn, dutta2019reachability, hu2020reach} process the overall closed-loop dynamics directly. The application of the recursive analysis framework to find an over-approximation of the reachable set in a finite horizon can be found in~\cite{tran2019safety, dutta2018learning, everett2021reachability, hu2020reach}. The effectiveness of the one-shot analysis in reducing compounding errors is demonstrated and analyzed in detail in OVERT~\cite{sidrane2022overt}~\footnote{The one-shot framework is denoted as the symbolic approach in~\cite{sidrane2022overt}.}, a safety verification method for nonlinear systems with NN controllers. In OVERT, the nonlinear dynamics is over-approximated by piecewise linear relations such that the safety of the closed-loop system can be verified through mixed-integer programming (MIP). Instead of focusing on one particular NN verification method, in this work we aim to provide guarantees on the performance of the one-shot framework for a range of NN verification methods including MIP~\cite{tjeng2017evaluating}, linear programming (LP)~\cite{wong2018provable}, and linear bounds propagation-based ones~\cite{zhang2018efficient, xu2020automatic}. Our analysis allows straightforward integration of these NN methods into the one-shot framework to further boost their performances in safety verification.

%Verification of neural network dynamical systems is closely related to open-loop neural network verification methods. Existing work combine output range analysis tools in reachability analysis frameworks, often resulting in recursive verification. For finite step reachable set over-approximation, a natural question to ask is what if we apply open-loop neural network verification to the unrolled closed-loop dynamics? Will it give tighter bounds?
%
%Although intuitive, we provide a counterexample showing the one-shot analysis may not necessarily lead to tighter bounds. This motivates our analysis in this work: provide formal guarantees on the tightness of the one-shot analysis. We show that it holds for many popular NN verification methods, including MILP, LP, backward linear bounds propagation. Furthermore, our analysis extends to general computational graphs, hence allowing uncertainty analysis in closed-loop. Based on the characterization of the gap between the recursive and one-shot approaches, we provide an adaptive bounding method in order to reduce the gap. 

\section{Preliminaries and Problem Formulation}
% formulation
\label{sec:formulation}
In this section, we formalize the reachability analysis problem for NNDS and two frameworks to approach it: the recursive and the one-shot analysis. Although, intuitively, the one-shot method should outperform the recursive method in tightness at the cost of computational overhead, we provide a motivating example that falsifies this claim, which calls for a formal comparison between these two frameworks.

\subsection{Neural network dynamical systems}
\begin{figure}[htb]
	\centering
	\includegraphics[width = 0.65 \columnwidth]{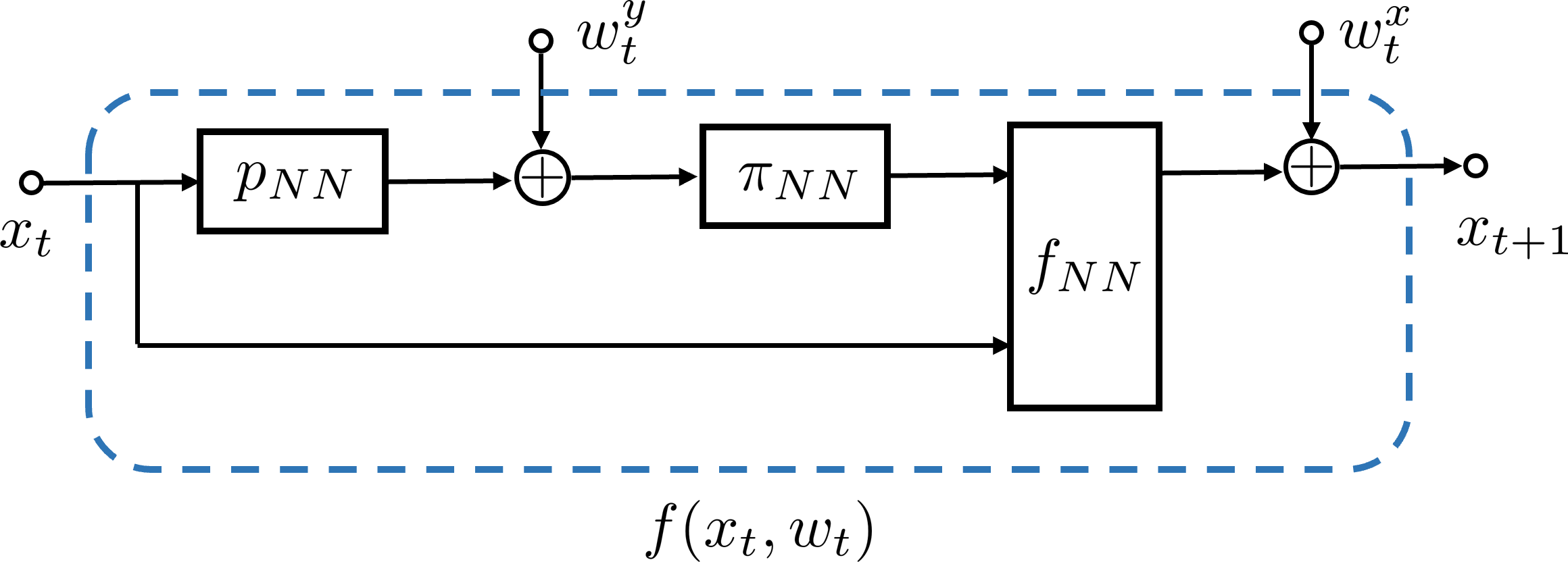}
	\caption{Illustration of a neural network dynamical system with several interconnected NNs.  }
	\label{fig:nnds}	
\end{figure}

We denote a discrete-time NNDS with disturbances as 
\begin{equation}\label{eq:nnds_w}
x_{t+1} = f(x_t, w_t)
\end{equation}
where $x_t \in \mathbb{R}^{n_x}$ is the state, $w_t \in \mathbb{R}^{n_w}$ denotes the disturbances at time $t$, and $f$ is a neural network with an arbitrary architecture taking $(x_t, w_t)$ as its input. For example, Eq.~\eqref{eq:nnds_w} can represent the closed-loop dynamics of a system consisting of several NN modules:
\begin{equation} \label{eq:nnds_connection}
\begin{cases}
x_{t+1} = f_{NN}(x_t, u_t) + w_t^x\\
y_t = p_{NN}(x_t) + w^y_t \\
u_t= \pi_{NN}(y_t)
\end{cases}
\end{equation}
where $f_{NN}, p_{NN}, \pi_{NN}$ denote, respectively, the NN dynamics, measurement function and control policies. The process and measurement disturbances $w_t^x$ and $w_t^y$ can be considered as components of a compound disturbance $w_t = [{w_t^x}^\top \ {w_t^y}^\top]^\top$. Then, $f(x_t, w_t)$ denotes a NN with the connection diagram shown in Fig.~\ref{fig:nnds}. 

\subsection{Finite-step reachability analysis}
For the NNDS~\eqref{eq:nnds_w}, define the one-step forward reachable set of $f(\cdot)$ from a given set $\mathcal{X} \times \mathcal{W} \subset \mathbb{R}^{n_x} \times \mathbb{R}^{n_w}$ as 
\begin{equation} \label{eq:output_set}
\begin{aligned}
f(\mathcal{X} \times \mathcal{W}) := \{y \in \mathbb{R}^{n_x} & \mid y = f(x, w), \  (x, w) \in \mathcal{X} \times \mathcal{W} \}.
\end{aligned}
\end{equation}
For a given set of initial conditions $\mathcal{X}_0 \subset \mathbb{R}^{n_x}$, the forward reachable set $\mathcal{R}_t(\mathcal{X}_0)$ of the NNDS~\eqref{eq:nnds_w} at time $t$ is defined by the recursion 
\begin{equation}
\begin{aligned}
\mathcal{R}_{t+1}(\mathcal{X}_0) = f(\mathcal{R}_t(\mathcal{X}_0) \times \mathcal{W}), \ \mathcal{R}_0(\mathcal{X}_0) = \mathcal{X}_0.
\end{aligned}
\end{equation}
Due to the complex structure of the function $f(\cdot)$, exactly identifying $\mathcal{R}_t(\mathcal{X}_0)$ is computationally challenging. Instead, we aim to find over-approximations $\bar{\mathcal{R}}_t(\mathcal{X}_0) \supseteq \mathcal{R}_t(\mathcal{X}_0)$ of the reachable sets at time $t$. Once these reachable set over-approximations are found, we can verify that the NNDS is safe, i.e., its trajectories starting from an initial set $\mathcal{X}_0$ never reach predefined unsafe regions in the state space over a finite horizon $T$ under disturbances. When the initial set is clear from context, we drop the $\mathcal{X}_0$ argument from $\overline{\mathcal{R}}_t$ for notational simplicity. 

%\begin{figure}[htb]
%	\centering
%	\includegraphics[width = 0.8 \columnwidth]{one_shot_framework}
%	\caption{In one-shot reachability analysis, the NN verification method is applied on the NNDS $f^{(T)}(\cdot)$ unrolled for $T$ steps directly to bound the reachable set of $x_T$. The unrolled NN takes the initial state $x_0$ and the disturbances $w_0, \cdots, w_{T-1}$ as the inputs, and $x_T$ as the output.}
%	\label{fig:one_shot_framework}
%\end{figure}

\subsection{Reachability analysis frameworks}
We first define the concept of \emph{propagator} as a basic reachability analysis module that abstracts different bounding methods of the output range of NNs.  
\begin{definition}[Propagator]
Any method $P$ that can bound the output of the NNDS $f(x, w)$ given a bounded input set $\mathcal{X} \times \mathcal{W}$ is defined as a \emph{propagator}, i.e., $P(\mathcal{X} \times \mathcal{W}; f) \subset \mathbb{R}^{n_x}$ and  $P(\mathcal{X} \times \mathcal{W}; f) \supseteq f(\mathcal{X} \times \mathcal{W})$.
\end{definition}

The first argument of the propagator. $\mathcal{X} \times \mathcal{W}$, denotes the input set while the second argument denotes the NN it acts on. The propagator can be obtained by running a NN verification algorithm, and its output set $P(\mathcal{X} \times \mathcal{W})$ often follows a pre-fixed geometric template specific to each NN verification algorithm, e.g., polytope~\cite{singh2019abstract}, ellipsoid~\cite{hu2020reach}, zonotope~\cite{singh2018fast}, or polynomial zonotope~\cite{kochdumper2022open}. In this work, we compare two frameworks that employ propagators to compute the reachable set over-approximations $\bar{\mathcal{R}}_t$ of the NNDS over a finite horizon:

\emph{a) Recursive reachability analysis} which applies the propagator $P$ for one-step reachable set over-approximation recursively, i.e., by using the reachable set over-approximation $\bar{\mathcal{R}}_t$ as the input set for computing $\bar{\mathcal{R}}_{t+1}$, as follows:

\begin{equation}\label{eq:recursive}
\begin{aligned}
&\textrm{\bf Recursive:} && \bar{\mathcal{R}}_0 = \mathcal{X}_0, \\
& && \bar{\mathcal{R}}_{t+1} = P(\bar{\mathcal{R}}_t \times \mathcal{W}; f), \ t \geq 0.
\end{aligned}
\end{equation}

\emph{b) One-shot reachability analysis} which applies the propagator on the $t$-th order composition of the NNDS~\eqref{eq:nnds_w} to compute $\bar{\mathcal{R}}_t$ directly for $t > 1$, as follows: 

\begin{equation} \label{eq:one_shot}
\begin{aligned}
&\textrm{\bf One-shot:} &&\bar{\mathcal{R}}_0 = \mathcal{X}_0, \\ &&&\bar{\mathcal{R}}_t = P(\mathcal{X}_0 \times \mathcal{W}^{t}; f^{(t-1)}), t \geq 1,
\end{aligned}
\end{equation}
where $f^{(t)}(x_0, w_{0:t})$ denotes the $t$-th order composition of the NNDS, i.e., $x_{t+1} = f^{(t)}(x_0, w_{0:t})$, and $\mathcal{W}^t$ denotes the Cartesian product of $t$ disturbance sets. The notation $w_{0:t}$ is shorthand for the set $\{w_0, \cdots, w_t\}$. Due to the cascading structure of NNs, one can easily represent the composition of the NNDS~\eqref{eq:nnds_w} over multiple time steps as a concatenated network. 

Intuitively, the one-shot method should generate tighter $\bar{\mathcal{R}}_t$ than the recursive method, since it utilizes the exact description $f^{(t)}(x_0, w_{0:t})$ of $x_{t+1}$ for reachable set over-approximation and incurs more computational cost due to the use of larger NNs ($t$ times the size of the NN considered in the recursive framework). However, we find a counterexample where the opposite holds, which shows that the one-shot verification method does not always result in bounds tighter than those obtained from the recursive framework.

\begin{figure}
	\centering
	\begin{subfigure}{0.45 \columnwidth}
		\centering 
		\includegraphics[width = \textwidth]{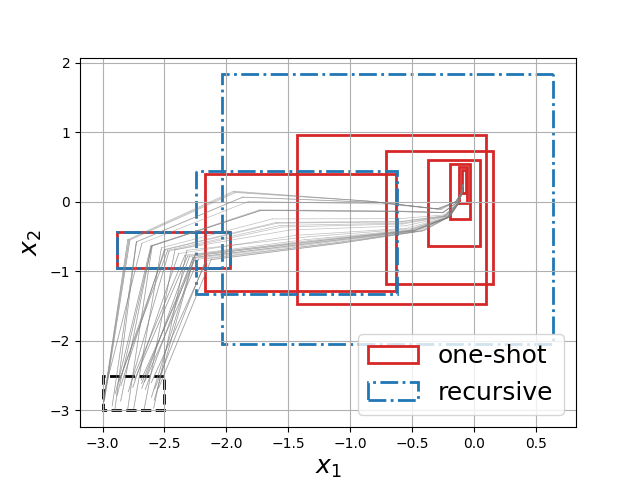}
		\caption{Backward bounds propagation.}
	\end{subfigure}
\hfil
	\begin{subfigure}{0.45 \columnwidth}
		\centering
		\includegraphics[width = \textwidth]{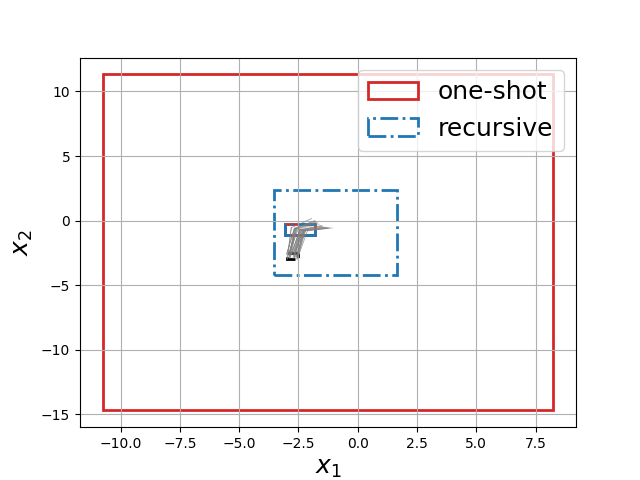}
		\caption{Forward bounds propagation.}
	\end{subfigure}
	\caption{Comparison of one-shot and recursive analysis of a NNDS approximating a Rayleigh-Duffing oscillator~\cite{gine2019dynamics}. Box reachable set over-approximations obtained by the backward (left)/forward (right) linear bounds propagation are plotted, as well as randomly sampled trajectories of the NNDS. In the left figure, the recursively synthesized bounds are shown for only $3$ steps. In the right figure, bounds over $2$ steps are plotted for both one-shot and recursive frameworks. 
	}
	\label{fig:counterexample}
\end{figure}

\begin{example}[Counterexample] \label{example:counterexample}
	We approximate the discretized dynamics of the 2D Rayleigh-Duffing oscillator~\cite{gine2019dynamics} as a feed-forward ReLU NN. No disturbances are considered. In Fig.~\ref{fig:counterexample}, the box reachable set over-approximations obtained from the one-shot and recursive frameworks are compared using the backward linear bounds propagation (left figure) and forward linear bounds propagation (right figure), respectively (see~\cite{xu2020automatic} for details). While the one-shot framework gives tighter bounds than the recursive one in the former case, the opposite holds in the latter.  
\end{example}

\subsection{Problem formulation}
Motivated by Example~\ref{example:counterexample}, in this paper we investigate the following question:

\begin{mdframed}
 In what cases is it beneficial to verify or optimize over the unrolled NN dynamics in one-shot?
\end{mdframed}

%\fbox{ In what cases is it beneficial to verify or optimize over the unrolled NN dynamics in one-shot? }

In Section~\ref{sec:analysis}, we characterize a general class of NN verification methods that are guaranteed to benefit from the one-shot analysis. Our proof applies to NNDS with arbitrary architecture. The implication of our analysis for verification algorithm design is discussed in Section~\ref{sec:applicability}, with numerical examples validating our analysis in Section~\ref{sec:simulation}.

\section{Tightness Improvement Guarantees}
% reachability analysis
\label{sec:analysis}
Neural networks implement mathematical functions that can be represented by a computational graph. In this section, we first describe the graph representation of NNs with arbitrary architecture, based on which the class of separable NN verification methods and propagators are defined. Then, a formal tightness improvement guarantee of the one-shot framework for separable propagators is provided in Theorem~\ref{thm:general_one_shot}. 

\subsection{NN representation}
Similar to~\cite{xu2020automatic}, we can represent a general NN through its computational graph, which is defined as a directed acyclic graph $\mathbf{G} = (\mathbf{V}, \mathbf{E})$, where $\mathbf{V} = \{z_0, \cdots, z_q, z_{q+1}, \cdots, z_L\}$. We denote $\mathcal{I} = \{z_0, \cdots, z_q\}$ as the input nodes, or independent nodes, of the graph and $\mathcal{D} =\{z_{q+1}, \cdots, z_L\}$ as the dependent nodes. The edge set $\mathbf{E}$ is a set of pairs $(i, j)$ denoting that node $z_i$ is an input argument of node $z_j$. Define $\text{Pre}(z_j) = \{z_{j,1}, \cdots, z_{j, m(j)}\}$ as the set of input nodes to $z_j$ with cardinality $\lvert \text{Pre}(z_j) \rvert = m(j)$. With this notation, we can describe the operator generating $z_j$ as $z_j = \phi_j(\text{Pre}(z_j))$ where $\phi_j$ can represent any commonly used operators in deep NNs such as linear, convolutional, MaxPooling, and activation layers, etc. We have $\lvert \text{Pre}(z_j) \rvert = 0$ for all input nodes $z_j \in \mathcal{I}$, and $\lvert \text{Pre}(z_j) \rvert > 0$ for the rest. All maps from the input nodes $\mathcal{I}$ to any dependent node $z_j \in \mathcal{D}$ are defined by the computational graph $\mathbf{G}$ and the operators $\phi_j(\cdot)$ associated to each node. We assume $z_L$ is the output of the NN whose property is of our concern. 

\textbf{Connection to NNDS}: Given $T > 0$, the unrolled NNDS $x_{T} = f^{(T)}(x_0, w_{0:T-1})$ for $T$ steps can be considered as a single NN with the computational graph $\mathbf{G}$. In this graph, the input nodes represent $\{x_0, w_0, \cdots, w_{T-1}\}$ and the dependent nodes represent all intermediate variables including states $x_{1:T-1}$. The state $x_T$ is the output node of $\mathbf{G}$, and our goal is to bound the reachable set of $x_T$ for safety verification.  

\subsection{Separable propagator}
In this paper, we consider NN verification methods that solve an optimization problem based on a layer-wise abstraction of $\phi_j(\cdot)$ in the NN computational graph. Specifically, these methods verify the properties of a NN by solving problems of the following form:

\begin{equation}\label{eq:general_verification}
	\begin{aligned}
		\underset{z_{0:L}}{\text{minimize}} & \quad J(z_L) \\
		\text{subject to} & \quad (\text{Pre}(z_k), z_k) \in \mathcal{S}_k, k = q+1, \cdots, L, \\
		& \quad z_k \in \mathcal{X}_k, k = 0, \cdots, q,
	\end{aligned}
\end{equation}
where $J$ is a real-valued function encoding the specification to be verified, $\mathcal{X}_k$ is the bounded input set of each input node $z_k$ for $0 \leq k \leq q$, and $\mathcal{S}_k$ denotes the constraints describing the relationship between node $z_k$ and its input nodes. We can view $\mathcal{S}_k$ as a set defined on the joint space of $(\text{Pre}(z_k), z_k)$, possibly described by expressions with intermediate variables, as shown in the following example. 

\begin{example}[Optimization constraints] 
\label{example:optimization}
Consider a scalar ReLU function $y = \max(x, 0)$. Given lower and upper bounds on the scalar input $x \in [\ell, u]$~\footnotemark, the constraint $(x,y) \in \mathcal{S}$ in different NN verification methods is formulated as: 
\footnotetext{We consider the non-trivial case where $\ell <0, u >0$. Otherwise, the ReLU layer is reduced to a linear one. }

a) Mixed-integer linear constraints~\cite{tjeng2017evaluating}:
\begin{equation}\label{eq:relu_MILP}
	y \geq 0,\ y \geq x,\ y \leq x - (1 - \mu) \ell,\ y \leq \mu u,\ \mu \in \{0, 1\}.
\end{equation}

b) Linear constraints~\cite{wong2018provable}:
\begin{equation} \label{eq:triangle_relaxation}
	y \geq 0,\ y \geq x,\ y \leq \frac{u}{u - \ell}x - \frac{u \ell}{u - \ell}.
\end{equation}

c) Simplified linear constraints~\cite{zhang2018efficient} that are amenable to linear bounds propagation:
\begin{equation}\label{eq:linear_bounds_relaxation}
	y \leq \frac{u}{u - \ell}x - \frac{u \ell}{u - \ell}, \  y \geq \alpha x.
\end{equation}
where $\alpha \in [0, 1]$ is a user-defined parameter. 

In~\eqref{eq:relu_MILP}, an intermediate binary variable $\mu$ is introduced to formulate $\mathcal{S}$ as the graph of the ReLU function. In~\eqref{eq:triangle_relaxation}, \eqref{eq:linear_bounds_relaxation}, the set $\mathcal{S}$ is a polyhedron in the space of $(x, y)$. The above abstractions of a ReLU function can be easily extended to all neurons in a NN, leading to the verification problem~\eqref{eq:general_verification}. 
\end{example}

If a NN verifier does not consider the coupling between nodes from different layers, the constraints $\mathcal{S}_k$ in problem~\eqref{eq:general_verification} are separable as defined below. 

\begin{definition}[Separable constraints] \label{def:separable_constr}
	Constraints	$\mathcal{S}_i$ and $\mathcal{S}_j$ such that $(x_i, y_i) \in \mathcal{S}_i$, $(x_j, y_j) \in \mathcal{S}_j$ are separable if the intermediate variables defining $\mathcal{S}_i$ and $\mathcal{S}_j$ are independent.
\end{definition}

\begin{definition}[Separable NN verifier]
	We denote any NN verification method that solves an optimization problem of the form~\eqref{eq:general_verification} with separable constraints $\mathcal{S}_k$ a separable NN verifier. 
%	Any propagator that applies a separable NN verifier is called a separable propagator.
\end{definition}

\paragraph{Generality of separable NN verifiers:} 
Separable NN verifiers include verification methods that either explicitly or \emph{implicitly} solve a primal optimization problem of the form~\eqref{eq:general_verification} with separable constraints. For example, for a ReLU network, all complete verifiers~\cite{katz2017reluplex, ehlers2017formal, bunel2018unified} are considered separable since they are equivalent to the MIP-based method~\cite{tjeng2017evaluating}. Similarly, scalable Lagrangian-based methods~\cite{bunel2020lagrangian, chen2022deepsplit} that solve the LP-based verification problem using the dual formulation are separable due to the strong duality of LP. Importantly, we observe that backward linear bounds propagation methods~\cite{zhang2018efficient, xu2020automatic}, which are scalable and can handle general activation functions, are also separable since they exactly solve an LP verification problem built on linear bounds abstractions of nonlinear activation functions with constraints similar to~\eqref{eq:linear_bounds_relaxation}. On the contrary, we cannot identify any implicit separable optimization problem for the forward linear bounds propagation method~\cite{xu2020automatic}. In fact, the forward method is used to construct the counterexample in Example~\ref{example:counterexample}.

\paragraph{Construction of the propagator:} 
A propagator can be constructed by applying a NN verifier to solve problem~\eqref{eq:general_verification} multiple times with different linear objective functions $J_i(z_L) = c_i^\top z_L, i = 1, \cdots, N$ in order to obtain a polytopic over-approximation of the output set of the NN. In what follows, we assume the normal vectors $\{ c_i \}_{i=1}^N$ defining the polytopic over-approximation are fixed. A propagator is called separable if it applies a separable NN verifier to bound the NN outputs. 

\subsection{Tightness comparison}
Recall that we can represent the $T$-step unrolled NNDS by a computational graph $\mathbf{G}$ with $x_T$ being the output node. In the one-shot framework, a propagator is directly applied to the computational graph $\mathbf{G}$. In the recursive framework, sub-networks of $\mathbf{G}$ are extracted such that the ranges of the intermediate nodes $x_{1:T-1}$ are over-approximated by the propagator as $\bar{\mathcal{R}}_t$ sequentially. Once the bound $\bar{\mathcal{R}}_t$ on the range of the intermediate state $x_t$ for $1 \leq t \leq T-1$ is computed, $x_t$ is treated as an input node to the downstream dependent nodes with input set $\bar{\mathcal{R}}_t$. We denote this process as \emph{concretization} of a dependent node in the computational graph $\mathbf{G}$.

We now show that for a separable propagator, concretizing the range of any dependent nodes in the NN computational graph $\mathbf{G}$ increases conservatism in bounding the output node. This immediately indicates that in reachability analysis with a separable propagator, the recursive framework is more conservative than the one-shot framework. Our analysis is conducted in two steps:

\paragraph{Recursive framework}
In the recursive framework, we first bound the range of a dependent node $z_k$ for some $q < k < L$. This can be achieved by applying a NN verifier solving problem~\eqref{eq:general_verification} with objectives $J_i(z_k) = c_i^\top z_k$, $i = 1, \cdots, N$ and the truncated constraints corresponding to the sub-network with $z_k$ as the output. To extract relevant constraints of this sub-network, we apply Breath First Search (BFS) on the graph $\mathbf{G}$ as shown in Algorithm~\ref{alg:BFS}. The set of extracted constraints is denoted as $\mathcal{S}_{input \rightarrow k} =$ \textsc{ExtractConstraints}(InputNodes = $z_{0:q}$, OutputNodes = $z_k$)~\footnote{We omit the arguments of the NN computational graph $\mathbf{G}$ and the constraint set $\{\mathcal{S}_i\}$ from~\eqref{eq:general_verification} since they are clear from context.}, and the NN verifier solves
\begin{equation}\label{eq:z_k_concretize}
	\begin{aligned}
		\text{minimize} & \quad J_i(z_k) \\
		\text{subject to} & \quad (z_{0:q}, z_k) \in \mathcal{S}_{input \rightarrow k}, \\
		& \quad z_{0:q} \in \mathcal{X},
	\end{aligned}
\end{equation}
for $i = 1, \cdots, N$ where the input set is given by $\mathcal{X} = \mathcal{X}_1 \times \cdots \times \mathcal{X}_q$. Denote $J_i^*$ the optimal values of~\eqref{eq:z_k_concretize}. Then, the over-approximation of the range of $z_k$ is given by~\footnote{Here we consider arbitrary dependent nodes $z_k$ in the computational graph of the unrolled NNDS. If the chosen node corresponds to state $x_t$, the over-approximation set $\bar{\mathcal{X}}_k$ from~\eqref{eq:z_k_set} coincides with $\bar{\mathcal{R}}_t$.  } 
\begin{equation} \label{eq:z_k_set}
	\bar{\mathcal{X}}_{k} = \{z \mid J_i(z) \geq J_i^*, i = 1, \cdots, N \}.
\end{equation}
Next, we treat $z_k$ as an input node with domain $ \bar{\mathcal{R}}_{k}$ to bound $z_L$. We consider the sub-network with $z_{0:1} \cup \{z_k \}$ as inputs and $z_L$ as output. We extract the associated constraints as $\mathcal{S}_{input, k \rightarrow L} =$ \textsc{ExtractConstraints}(InputNodes = $z_{0:q} \cup \{z_k\}$, OutputNodes = $z_L$). Bounding the range of $z_L$ is now achieved by solving
\begin{equation}\label{eq:z_L_concretize}
	\begin{aligned}
		\text{minimize} & \quad J_i(z_L) \\
		\text{subject to} & \quad (z_{0:q}\cup z_k, z_L) \in \mathcal{S}_{input,k \rightarrow L}, \\
		& \quad z_{0:q} \in \mathcal{X}, z_k \in \bar{\mathcal{X}}_k,
	\end{aligned}
\end{equation}
for $i = 1, \cdots, N$. Denote the optimal values of~\eqref{eq:z_L_concretize} as $J_{i, re}^*$. The over-approximation of $z_L$ obtained by the recursive framework is given by $\bar{\mathcal{X}}_{L, re} = \{ z \mid J_i(z) \geq J_{i,re}^*, 1 \leq i \leq N\}$. 

\paragraph{One-shot framework}
In the one-shot analysis framework, we directly solve problem~\eqref{eq:general_verification} with objectives $J_i(z_L)$ for $i =1, \cdots, N$. Denote the optimal values of these problems as $J_{i, os}^*$. Then, the over-approximation of $z_L$ obtained by the one-shot framework is given by $\bar{\mathcal{X}}_{L, os} = \{z \mid J_i(z) \geq J_{i, os}^*, 1 \leq i \leq N \}$.

\begin{algorithm}[htb]
	\caption{Extract concretization constraints}\label{alg:BFS}
	\hspace*{\algorithmicindent} \textbf{Input:} InputNodes $\subseteq \{z_i\}_{i=0}^{L}$, OutputNodes $z_k$, computational graph $\mathbf{G}$, constraints $\{\mathcal{S}_i\}_{i=q+1}^L$ from problem~\eqref{eq:general_verification}. \\
	\hspace*{\algorithmicindent} \textbf{Output:} Constraint set $\mathcal{S}$.  
	\begin{algorithmic}[1]
		\Procedure{ExtractConstraints}{}
		\State Let $Q$ be a queue and $\mathcal{S} = \emptyset$.
		\State $Q.\text{enqueue}(z_k)$. Mark node $z_k$ as explored. 
		\While {$Q$ is not empty}
		\State $z_i$ = $Q.\text{dequeue}()$. \Comment{Subscript $i$ denotes the label of the node in graph $\mathbf{G}$.}
		\State $\mathcal{S} = \mathcal{S} \cup \mathcal{S}_i$.
		\For {$z_j$ in $\text{Pre}(z_i)$}
		\If {$z_j \notin$ InputNodes and is not explored}
		\State $Q.enqueue(z_j)$ and label $z_j$ as explored.
		\EndIf
		\EndFor
		\EndWhile
		\EndProcedure
	\end{algorithmic}
\end{algorithm}

\begin{lemma} \label{lem:general_concretization}
With a separable propagator, the one-shot framework generates a tighter over-approximation of the NN output than the recursive framework, i.e., $\bar{\mathcal{X}}_{L, os} \subseteq \bar{\mathcal{X}}_{L, re}$. 
\end{lemma}

\begin{proof}
To allow a straightforward comparison, we combine the two-step computation~\eqref{eq:z_k_concretize} and~\eqref{eq:z_L_concretize} of the recursive framework into one optimization problem by introducing auxiliary variables $\tilde{z}_{0:L}$ and $\tilde{z}^j_{0:L}$ for $j = 1, \cdots, N$. We can equivalently rewrite~\eqref{eq:z_L_concretize} as

\begin{equation}\label{eq:general_recursive}
\begin{aligned}
\text{minimize} & \quad J_i(\tilde{z}_L) \\
\text{subject to} & \quad (\tilde{z}_{0:q}^j, \tilde{z}_k^j) \in \mathcal{S}_{input \rightarrow k}, \ \tilde{z}_{0:q}^j \in \mathcal{X}, \\
& \quad J_i(\tilde{z}_k) \geq J_i(\tilde{z}_k^j), j = 1, \cdots, N, \\
& \quad (\tilde{z}_{0:q} \cup \tilde{z}_k, \tilde{z}_L) \in \mathcal{S}_{input, k \rightarrow L}, \\
& \quad \tilde{z}_{0:q} \in \mathcal{X}.
\end{aligned}
\end{equation}
By the separability of constraints $\mathcal{S}_k$ in~\eqref{eq:general_verification}, we have that any feasible solution $z_{0:L}$ of~\eqref{eq:general_verification} constructs a feasible one of~\eqref{eq:general_recursive} by setting $\tilde{z}_\ell = \tilde{z}^j_\ell = z_\ell$ for all $0 \leq \ell \leq L$, $1 \leq j \leq N$ with the same objective value. Then we have $J_{i, os}^* \geq J_{i, re}^*$ for all $1 \leq i \leq N$ and $\bar{\mathcal{X}}_{L, os} \subseteq \bar{\mathcal{X}}_{L, re}$.
\end{proof}

%\begin{remark}
%	Lemma~\ref{lem:general_concretization} can be extended to the case when more than one dependent nodes is concretized. The grouping of constraints $\mathcal{S}_k$ with respect to the chosen dependent nodes and the proof of the tightness gap follow similarly.
%\end{remark}

The proof of Lemma~\ref{lem:general_concretization} shows that the performance gap between the one-shot and recursive frameworks arises from the concretization errors, i.e., the over-approximation error of applying a finite number of hyperplanes to bound the reachable set of a dependent node in the NN. Such errors may compound when we concretize the ranges of a sequence of dependent nodes one-by-one as in recursive reachability analysis of the NNDS~\eqref{eq:nnds_w}.

\begin{theorem} \label{thm:general_one_shot}
The one-shot framework is guaranteed to generate tighter over-approximations of reachable sets than the recursive framework for the NNDS~\eqref{eq:nnds_w} when a separable propagator is applied. 
\end{theorem}
\begin{proof}
	We can treat the $T$-step unrolled NNDS as the NN with input nodes $(x_0, w_0, \cdots, w_{T-1})$ and $x_T$ as the output node. By Lemma~\ref{lem:general_concretization}, concretizing any intermediate dependent node $x_t$ leads to more conservative over-approximation of $x_T$. 
\end{proof}

\section{Practical Implications}
% discussion on the applicability of our analysis
\label{sec:applicability}
We now summarize the comparison between the one-shot and recursive frameworks: (a) The one-shot framework solves the NN verification problem on NNs $f^{(t)}(\cdot)$ with growing sizes as $t$ increases, while in the recursive framework a NN $f(\cdot)$ of constant size is considered. Therefore, the one-shot framework is always more computationally costly to run than the recursive framework. (b) The recursive framework generates looser bounds than the one-shot framework due to the concretization error as shown in Lemma~\ref{lem:general_concretization} and the induced conservatism could explode as time $t$ increases. 

\paragraph{One-shot analysis principle:} 
The above two points indicate that we can treat the size of NNs, or the order of composition of the NNDS $f(\cdot)$, as a tuning parameter which makes a trade-off between tightness and computational complexity. Within a given computational resource budget (either in time or in memory), our analysis suggests that we should run a separable propagator on the composition $f^{(t)}$ of the NNDS with $t$ as large as possible. 

\begin{figure}
	\centering
	\begin{subfigure}{0.45 \columnwidth}
		\includegraphics[width = 0.99 \textwidth]{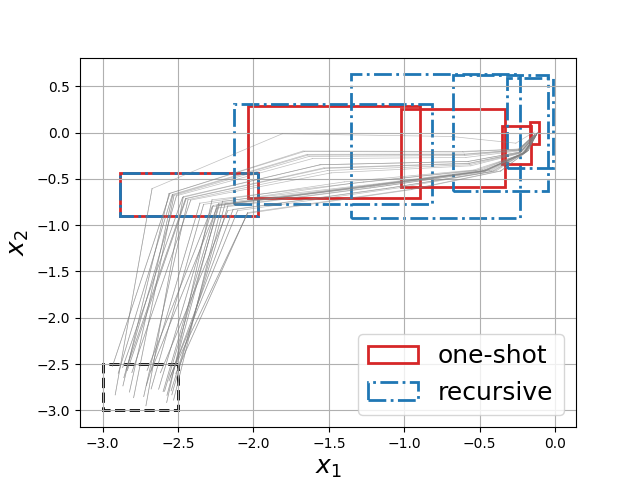}
		%		\caption{Box bounds with $4$ normal vectors. }
		\label{fig:duffing_LP_box}
	\end{subfigure}
\hfil
	\begin{subfigure}{0.45 \columnwidth}
		\includegraphics[width = 0.99 \textwidth]{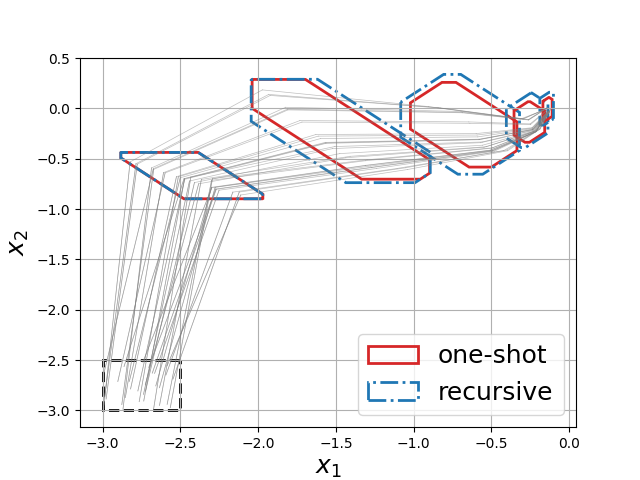}
		%		\caption{Polyhedral bounds with $8$ normal vectors. }
		\label{fig:duffing_LP_poly}
	\end{subfigure}
	\caption{The LP-based propagator using relaxation~\eqref{eq:triangle_relaxation} is applied to bound the reachable sets of the Rayleigh-Duffing oscillator NNDS using a box template (left, $4$ hyperplanes) and a more complex polyhedral template (right, $8$ hyperplanes). Randomly sampled trajectories of the NNDS are shown in gray. The tightness gap between the one-shot and recursive frameworks decreases as the complexity of the bounding templates increases. }
	\label{fig:duffing_LP}
\end{figure}

\paragraph{Concretization error reduction:} 
Problem~\eqref{eq:general_recursive} indicates that increasing $N$ tightens the search space of $\tilde{z}_k$ and consequently the output variable $\tilde{z}_L$. Therefore, the concretization error from the recursive framework can be decreased by considering a richer set of objective functions $\{c_i^\top z_L\}$ in constructing the propagator. This phenomenon is demonstrated in Fig.~\ref{fig:duffing_LP}. However, the overall complexity of the recursive framework increases with $N$, and it remains a challenge to identify a set of effective bounding hyperplanes $\{c_i \}_{i=1}^N$ for systems with high dimensions.

\section{Numerical Examples}
% simulation
\label{sec:simulation}
We demonstrate the improvement of tightness of the one-shot analysis on a NNDS that approximates the closed-loop dynamics of a cart-pole system. Separable propagators relying on LP and backward linear bounds propagation are considered with different NNDS structures. All experiments are run on an Intel Core i7-6700K CPU.

\subsection{LP-based propagator}
\begin{figure}
	\centering
	\begin{subfigure}{0.45 \columnwidth}
		\includegraphics[width = 0.99 \textwidth]{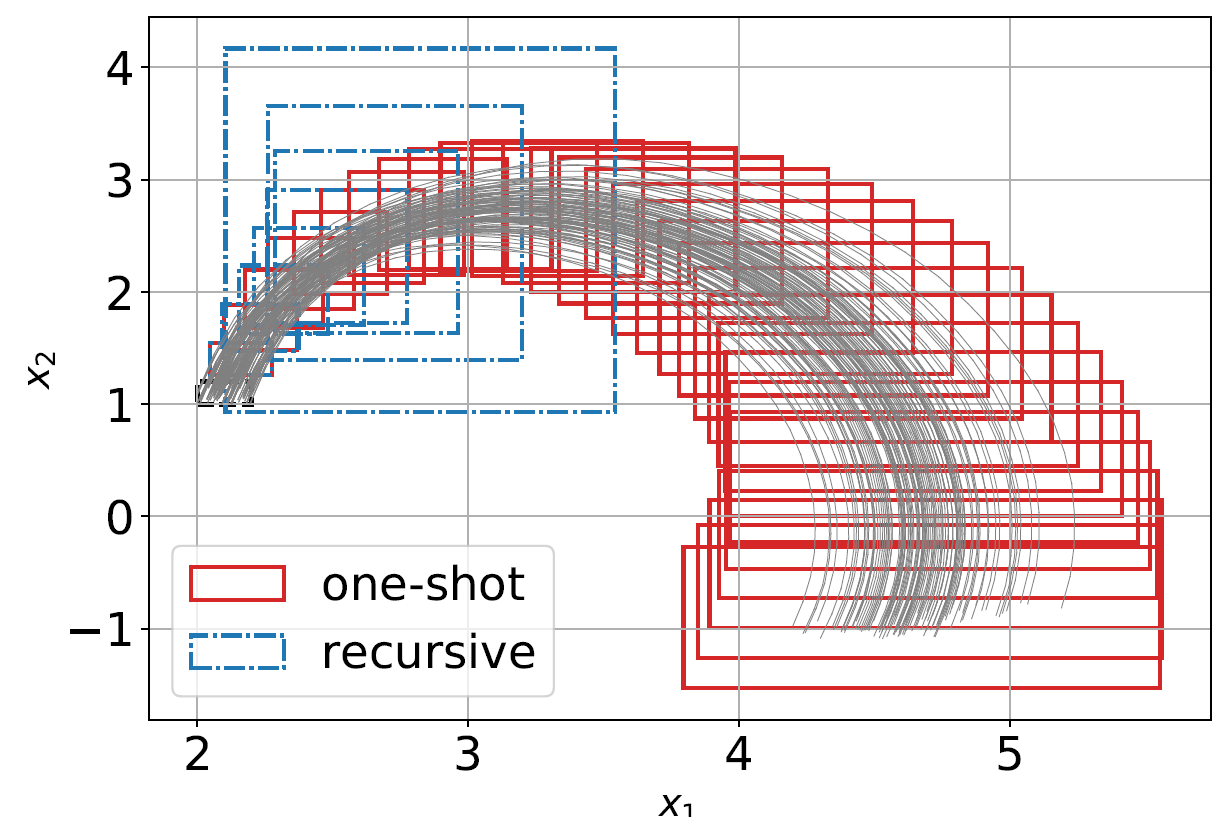}
		%		\caption{Box bounds with $4$ normal vectors. }
		\label{fig:x12_comparison}
	\end{subfigure}
\hfil
	\begin{subfigure}{0.45 \columnwidth}
		\includegraphics[width = 0.99 \textwidth]{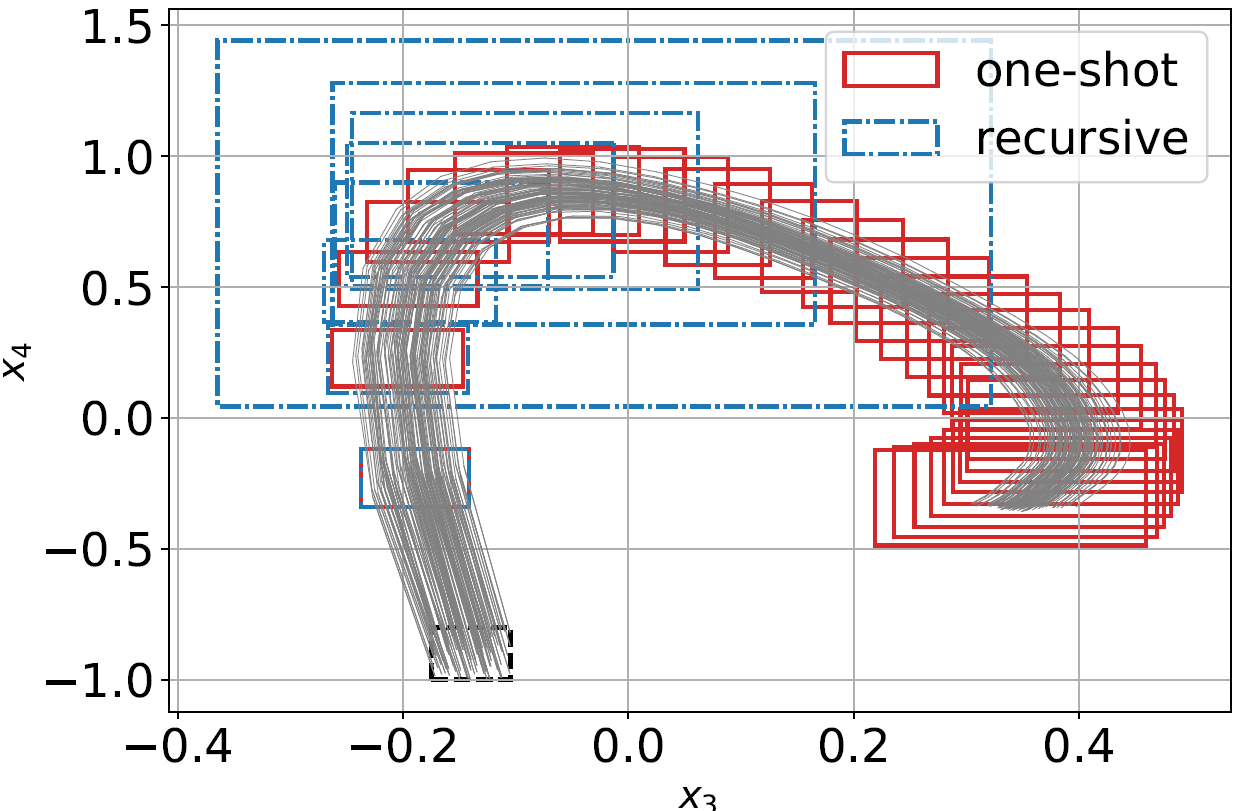}
		%		\caption{Polyhedral bounds with $8$ normal vectors. }
		\label{fig:x34_comparison}
	\end{subfigure}
	\caption{
 Box reachable set over-approximations of the feedforward NNDS~\eqref{eq:cartpole_forward} obtained by the one-shot (solid red boxes, $30$ steps) and recursive (dashed blue boxes, $8$ steps) frameworks using the LP-based propagator. The boxes are projected onto the $(x_1,x_2)$ and $(x_3, x_4)$ planes for plotting. Randomly sampled trajectories of the NNDS are shown in gray.
 }
	\label{fig:sim_comparison}
\end{figure}

We consider the cart-pole dynamics without friction from~\cite{florian2007correct} where the cart has a mass of $0.25$ kg, and the pole has a mass of $0.1$ kg and length of $0.4$ m. The entries of the $4$-dimensional state $x = [x_1 \ x_2 \ x_3 \ x_4]^\top$ represent the position, the velocity of the cart, the angle, and the angular speed of the pole, respectively. We train a feedforward ReLU NN $f_{cl, NN}(x)$ with the structure $4 - 100 - 100 - 4$ to approximate the discretized closed-loop dynamics of the cart-pole system under a nonlinear model predictive controller (MPC)~\cite{lucia2017rapid} with $dt = 0.05$s. An LP-based verifier that bounds ReLU by~\eqref{eq:triangle_relaxation} is applied to compute reachable sets of the NNDS 
\begin{equation}\label{eq:cartpole_forward}
x_{t+1} = f_{cl, NN}(x_t).
\end{equation}
The pre-activation bounds $(\ell, u)$ in~\eqref{eq:triangle_relaxation} are computed in an inductive manner by the verifier. The initial set is given by $\mathcal{X}_0 = \{x\mid x_1 \in [2.0, 2.2] \textrm{ m}, x_2 \in [1.0, 1.2] \textrm{ m/s}, x_3 \in [-0.174, -0.104] \textrm{ rad}, x_4 \in [-1.0, -0.8] \textrm{ rad/s} \}$. In Fig.~\ref{fig:sim_comparison}, the over-approximating boxes obtained by the recursive method (dashed blue) for $8$ steps are plotted which quickly become overly conservative, while the one-shot method is able to provide informative bounds (solid red) on the system states for $30$ steps. The LP propagators are solved by Gurobi~\cite{gurobi}, and the total solver time for running the one-shot and recursive methods for $8$ steps are $161.3$ and $3.1$ seconds, respectively.

\subsection{Backward propagation-based propagator}
\begin{figure}
	\centering
	\begin{subfigure}{0.45 \columnwidth}
		\includegraphics[width = 0.99 \textwidth]{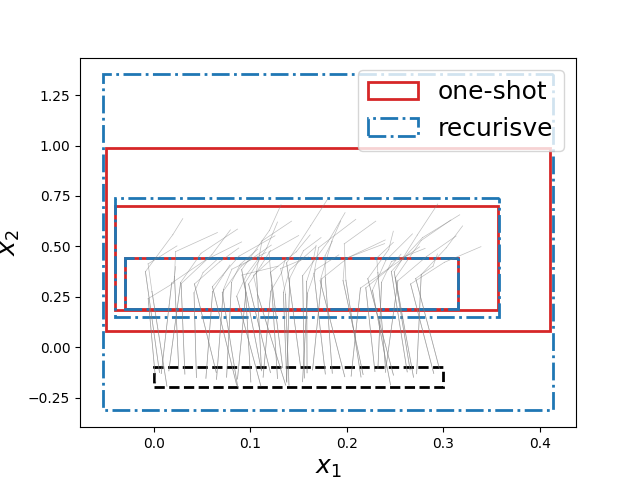}
		%		\caption{Box bounds with $4$ normal vectors. }
		\label{fig:complex_x12}
	\end{subfigure}
\hfil
	\begin{subfigure}{0.45 \columnwidth}
		\includegraphics[width = 0.99 \textwidth]{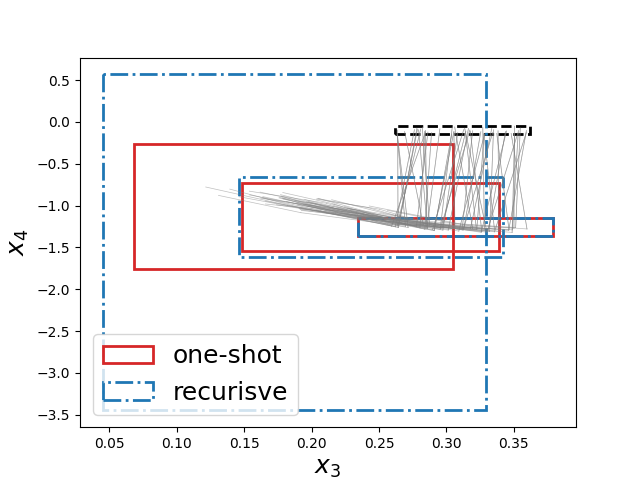}
		%		\caption{Polyhedral bounds with $8$ normal vectors. }
		\label{fig:complex_x34}
	\end{subfigure}
	\caption{Box reachable set over-approximations of the closed-loop NNDS~\eqref{eq:complex} obtained by backward linear bounds propagation are plotted for $3$ steps. Bounds produced by the one-shot framework (red boxes) are tighter than those from the recursive framework (blue boxes). Randomly sampled trajectories are plotted in gray. }
	\label{fig:complex}
\end{figure}

Next, we adopt a different approximation of the closed-loop dynamics of the cart-pole system which has a more complex structure:
\begin{subequations}\label{eq:complex}
\begin{align}
\label{eq:1}
&x_{t+1} = Ax_t + B u_t + f(x_t, u_t) + w_t \\
\label{eq:2}
&u_t= \pi_{NN}(x_t)
\end{align}
\end{subequations}
where the linear dynamics $(A, B)$ are derived from the Jacobian matrices of the cart-pole dynamics at the origin. A feedforward NN $f(x, u): \mathbb{R}^5 \mapsto \mathbb{R}^4$ is trained to learn the residual nonlinear dynamics with training data randomly sampled from the discretized cart-pole dynamics with $dt = 0.05$s. We implement a nonlinear MPC for the open-loop NNDS~\eqref{eq:1} in MATLAB to collect samples of $(x_t, u_t)$ and train the NN controller $\pi(\cdot): \mathbb{R}^4 \mapsto \mathbb{R}$ in PyTorch to approximate the MPC policy. The residual NN dynamics $f(x,u)$ in~\eqref{eq:1} is a feedforward tanh network of structure $5-50-4$, and the controller $\pi(x)$ in~\eqref{eq:2} is a feedforward ReLU network of structure $4-100-1$. The disturbance $w_t$ is assumed bounded by $\lVert w_t \rVert_\infty \leq 0.05$. We compare the one-shot and recursive frameworks on system~\eqref{eq:complex} using backward linear bounds propagation~\cite{xu2020automatic} since it can handle general activation functions. The results are shown in Fig.~\ref{fig:complex} where the initial set is chosen as $\mathcal{X}_0 = \{x\mid x_1 \in [0.0, 0.3] \textrm{ m}, x_2 \in [-0.2, -0.1] \textrm{ m/s}, x_3 \in [0.262, 0.312] \textrm{ rad}, x_4 \in [-0.15, -0.05] \textrm{ rad/s} \}$ and the horizon is set as $T = 3$~\footnotemark. We observe that the one-shot framework generates tighter bounds than the recursive framework, validating Theorem~\ref{thm:general_one_shot}. The total solver time is $2.22$s for the one-shot framework and $1.01$s for the recursive framework.

\footnotetext{Backward linear bounds propagation solves a weaker relaxation of the NN verification problem than LP-based methods. Therefore, in this example, we only consider $T = 3$ since both the one-shot and recursive frameworks become too conservative afterward.}

\section{Conclusion}
% conclusion
\label{sec:conclusion}
We proved that for a wide range of NN verification methods, analyzing the unrolled dynamics of a NN dynamical system generates tighter bounds than a recursive, step-by-step analysis. Our proof holds for NN dynamical systems with general computational graphs. The applicability of one-shot reachability analysis was demonstrated numerically. Future work includes designing adaptive bounding polytopes such that the performance gap between the recursive and one-shot frameworks is minimized.

\section*{Acknowledgment}
We thank Chelsea Sidrane for useful discussions. 

%\addtolength{\textheight}{-12cm}   

\bibliographystyle{ieeetr}
\bibliography{Refs}

\end{document}